\documentstyle[11pt,epsf]{article}

\newcommand{\bee}{\begin{equation}}
\newcommand{\ee}{\end{equation}}
\newcommand{\beea}{\begin{eqnarray}}
\newcommand{\eea}{\end{eqnarray}}

\newcommand{\rmd}{{\rm d}}

\def\nabstar#1{\nabla\kern-0.5pt\smash{\raise 4.5pt\hbox{$\ast$}}
               \kern-4.5pt_{#1}}

\def\RPn{{\rm RP}^{n-1}}
\def\RP2{{\rm RP}^2}
\def\On{{\rm O}(n)}
\def\O3{{\rm O}(3)}
\def\SUn{{\rm SU}(n)}

\def\SV{{\bf S}}
\def\sig{{\bf \sigma}}
\def\A{{\cal A}}
\def\L{{\cal L}}
\def\CP{{\cal P}}

\begin{document}
\thispagestyle{empty}
\parskip=12pt
\raggedbottom

\def\mytoday#1{{ } \ifcase\month \or
 January\or February\or March\or April\or May\or June\or
 July\or August\or September\or October\or November\or December\fi
 \space \number\year}
\noindent
\hspace*{9cm} MPI--PhT/95--55\\
\hspace*{9cm} BUTP--95/19\\
\vspace*{1cm}
\begin{center}
{\LARGE On the question of universality in \\
$\RPn$ and $\On$ Lattice Sigma Models\footnote{Work supported in part by Schweizerischer Nationalfonds}}

\vspace{1cm}

Ferenc Niedermayer\footnote{On leave from the Institute of Theoretical
Physics, E\"otv\"os University, Budapest}
\\
\vskip 1ex
Institute for Theoretical Physics \\
University of Bern \\
Sidlerstrasse 5, CH-3012 Bern, Switzerland

\vspace{0.5cm}

Peter Weisz and Dong-Shin Shin\\
\vskip 1ex
Max-Planck-Institut f\"ur Physik\\
F\"ohringer Ring 6, D-80805 M\"unchen, Germany \\

July 1995 \\ \vspace*{0.5cm}

\nopagebreak[4]

\begin{abstract}
We argue that there is no essential violation of universality in the continuum
limit of mixed $\RPn$ and $\On$ lattice sigma models in 2 dimensions, contrary
to opposite claims in the literature.
\end{abstract}

\end{center}
\eject

\section{Introduction}

In this paper we consider two--dimensional mixed isovector-isotensor
$\On$ sigma models described by a lattice action of the kind
\bee
\A(S)=\beta_V \sum_{x,\mu} \left( 1-\SV_x \SV_{x+\mu} \right)
+{1\over 2} \beta_T \sum_{x,\mu} \left(1-(\SV_x \SV_{x+\mu})^2\right),
\label{1}
\ee
with $\SV_x^2=1$. The sums run over the nearest neighbor sites.
This provides a possible lattice discretization for the continuum
$\On$ non--linear sigma model,
\bee
\A^{cont}={1 \over 2} \beta \int \rmd^2 x \left(\partial_{\mu} \SV(x)
\right)^2
\label{2}
\ee
with $\beta=\beta_V+\beta_T$.
 
According to conventional wisdom, different lattice regularizations
(preserving the crucial symmetries)
yield the same continuum field theory (``universality'').
For the case of the action (\ref{1}),
Caracciolo, Edwards, Pelissetto and Sokal \cite{CEPSI,CEPSII},
however, question this assumption
and in particular state that the pure sigma model ($\beta_T=0$)
and the pure $\RPn$ model ($\beta_V=0$) have different
continuum limits for $\beta \to \infty$.
Since the notion of universality plays an essential role in the
theory of critical phenomena it is worthwhile to consider this
question again.
In this paper we will explain
how the peculiar features observed
in the model (\ref{1}) can be understood in the framework
of the conventional picture. We wish to stress, however,
that our scenario is (for the most part)
based on plausibility arguments, for which
rigorous proofs are unfortunately still lacking.

A related problem concerns the
mixed fundamental--adjoint action in
pure $\SUn$ gauge theory \cite{mixed} in 4 dimensions.
The generally accepted
belief is that there is a universal continuum limit for these
theories. However, we shall not discuss this model here.
 
The paper is organized as follows.
In section 2 we consider a class of pure $\RPn$ models. We first
describe some general properties and then
go on to discuss the continuum limit.
Section 3 presents an investigation of
perturbed $\RPn$ models, paying
special attention to their expected continuum limit.
In particular, we argue there is
no contradiction to the general understanding of universality.
Finally in section 4 we outline some calculations supporting
our general scenario.

\section{The $\RPn$ models}
 
\subsection{Some general properties}

The standard action of the $\RPn$ model is
\bee
\A_T(S)=
{1\over 2} \beta \sum_{x,\mu}\left(1-(\SV_x \SV_{x+\mu})^2\right).
\label{3}
\ee
It has, compared with the $\On$ model, an extra local $Z_2$
symmetry: it is invariant under the transformation
\bee
\SV_x \to g_x \SV_x, \,\,{\rm where} \,\, g_x=\pm 1.
\label{4}
\ee
As a consequence, only those quantities have non--zero expectation
values which are invariant under this local transformation.
In particular the isovector correlation function vanishes:
\bee
\langle \SV_x \SV_y \rangle =0 ~~~~{\rm for} ~~~~ x \ne y.
\label{5}
\ee
The simplest local operator with non--vanishing correlation function
is the tensor
$T_x^{\alpha \beta}
= \SV_x^{\alpha} \SV_x^{\beta}-\delta^{\alpha \beta}/n$:
\bee
\langle T_x^{\alpha \beta} T_y^{\alpha \beta} \rangle \neq 0.
\label{6}
\ee

This behavior seems completely different from
that of the $\On$ sigma model, so that
one might expect drastic differences in the physics described
by the models. This is indeed true for the theories with finite
lattice spacing, but
below we shall argue that in the  continuum limit
this difference becomes insignificant, and can be resolved
by consideration of nonlocal variables.
 
\subsection{Defects and phase structure}

For convenience, we introduce the notation
$u_{xy} \equiv \SV_x \SV_y$
for the scalar product of two spins.
Further, for any path $\CP$ on the lattice define the observable
\bee
W(\CP)=\prod_{ <x,y> \in \CP } u_{xy} \,,
\label{7}
\ee
where $<x,y>$ denotes the link joining two neighboring points
$x$ and $y$.

Consider a configuration of the $\RPn$ model. One says that it
has a defect associated with a plaquette $p$
(or a site on the dual lattice) if
\bee
W(\partial p) < 0 \,,
\label{8}
\ee
where $\partial p$ is the boundary of the plaquette.
The defects are endpoints of paths
on the dual lattice formed by those dual links with $u_{xy} < 0$,
where $x,y$ are the two sites on the corresponding link.
Due to the local gauge invariance, only the position of the defects
is physical, while the paths can be moved by a gauge transformation.

Like the vortices in the two-dimensional XY model \cite{KoTh},
these defects play
an essential role in determining the phase structure of the
$\RPn$ model at finite $\beta$ \cite{Wolff}.
Some of these aspects are discussed
by Kunz and Zumbach \cite{KZII}.
The activation energy of a pair of defects grows logarithmically 
with their separation $r$.
The standard energy--entropy argument \cite{KoTh} then predicts
a phase transition at some finite $\beta_c$. For $\beta<\beta_c$ the
defects are deconfined, while for $\beta>\beta_c$ they appear in
closely bound pairs.
This difference is expected to show up in an area or perimeter law
(for $\beta<\beta_c$ and $\beta>\beta_c$ respectively)
of the ``Wilson loop'' expectation value
$\langle W(\L) \rangle$ for large loops $\L$ \cite{KZII}.

We see this in a large $n$ limit
of the $\RPn$ model \cite{MR,KZI}.
There the phase transition is demonstrated to be first order.
Furthermore, one verifies the expected ``Wilson loop'' signal:
in the leading order,
$\langle W(\L) \rangle = 0$ for $\beta<\beta_c$, while
$\langle W(\L) \rangle = \exp \{ -\gamma(\beta) |\L| \}$
for $\beta>\beta_c$, with $|\L|$ the perimeter of $\L$.

For finite $n$, however, the situation is not at all clear.
The discussion of the nature of the critical point at finite $\beta$
has a long history \cite{Sol,Fuk,Sinc,KZII,CEPSII}.
All MC simulations show that approaching $\beta_c$ from below
the correlation length starts to grow drastically. 
However, the various authors disagree concerning the nature of
the transition,
the variety of opinions based merely on theoretical expectations
(and prejudices).
We shall return to this  question later.

In the following we will discuss
the possible continuum limits.
We shall argue that at finite $\beta$ the
correlation length in the $\RPn$ model always stays finite,
and the critical point at $\beta=\infty$ is equivalent to that 
of the $\On$ model.

\subsection{Equivalence of the $\RPn$ and $\On$ models
in the continuum limit}

Consider a more general form of the lattice $\RPn$ action:
\bee
\A_T(S)=\beta \sum_{<x,y>} f\left( u_{xy} \right),
\label{9}
\ee
where the function $f(u)$ satisfies the following properties:
\bee
f(-u)=f(u),~~f(1)=0,~~f'(1)=-1
\label{10}
\ee
and $f(u)$ is monotonically decreasing for $0<u<1$.
We assume a weaker form of universality: any of these choices
yields the same continuum limit as $\beta\to\infty$.
(Actually, even less will be sufficient --- one can keep fixed
the form of $f(u)$ for $u_0<|u|<1$ to be the standard one.)

Let us now introduce a chemical potential $\mu$ of the defects
modifying the Boltzmann factor by $\exp(-\mu n_{def})$
where $n_{def}$ is the number of defects. At $\mu>0$ the defects 
are suppressed and at $\mu=\infty$ no defects are allowed.

Take first the $\mu=\infty$ case.
As was done by Patrascioiu and Seiler \cite{PS},
one can define Ising variables $\epsilon_x=\pm 1$ by
\bee
\epsilon_x = {\rm sign} \{ W(\CP_{x_0 x})\} \,,
\label{11}
\ee
starting from a fixed site $x_0$ and going along some path $\CP_{x_0 x}$
connecting $x_0$ to $x$.
Due to the absence of defects, $\epsilon_x$ does not
depend on the path chosen. For two nearest neighbor sites one has
\bee
\epsilon_x \epsilon_{x+\mu} ={\rm sign}(u_{x x+\mu}).
\label{12}
\ee
Introduce now a new $\On$ vector
\bee
\sig_x = \epsilon_x \SV_x.
\label{13}
\ee
This has the property that $\sig_x \sig_{x+\mu} = |\SV_x \SV_{x+\mu}| >0$
for nearest neighbors.
The dynamics of the $\sig_x$ field is described by the modified
$\On$ action
\bee
\A_V(\sig)=\beta \sum_{x,\mu} f_V\left( \sig_x \sig_{x+\mu}\right)
\label{14}
\ee
with
\bee
f_V(u)=  \cases{f(u) \quad {\rm for} \,\, u\ge 0 \,, \cr
         \noalign{\vskip2pt}
         \infty \quad \quad {\rm for} \,\, u<0 \,. \cr}
\label{15}
\ee
We also assume that the continuum limit ($\beta\to\infty$)
for this action is the same as for the standard $\On$ action
(universality {\it within} the $\On$ model).

The $\RPn$ model described by (\ref{9})
at $\mu=\infty$ and the corresponding $\On$ model given by (\ref{14})
are equivalent in the continuum limit in the following sense:
all gauge invariant quantities (such
as the tensor correlation function or a Wilson loop of scalar products)
in the $\RPn$ model are exactly the same as in the
$\On$ model, while all non-gauge invariant quantities
vanish in the $\RPn$ model. In particular, for
the vector correlation function 
\bee
\langle \SV_x \SV_y \rangle = \langle \epsilon_x \epsilon_y \rangle
\langle \sig_x \sig_y \rangle = 0 \,\, {\rm for} \,\, x\ne y
\label{16}
\ee
since $\langle \epsilon_x \epsilon_y \rangle = \delta_{xy}$.
The $\SV_x$ vector of the $\RPn$ model can be thought of
as a product of two independent fields,
the ``true vector'' $\sig_x$ and the Ising variable $\epsilon_x$;
one is described by the corresponding $\On$ model,
while the other by an Ising model at infinite temperature. 

We return now to the case of $\RPn$ model at finite $\mu$.
With increasing $\mu$ the average defect density is decreased.
Defects tend to disorder the system, therefore
it is very plausible to assume that the correlation
length (in the tensor channel) grows with increasing $\mu$.
Since at $\mu=\infty$ the $\RPn$ model is equivalent
to the corresponding $\On$ model at the same $\beta$,
one concludes that the correlation length
at $\mu=0$ is bounded by that of the $\On$ model.

Assuming further that, according to the standard scenario,
the $\On$ model has a finite correlation length for finite $\beta$,
it follows that the $\RPn$ model cannot have a phase transition
(at finite $\beta$) with diverging correlation length.

The latter is in agreement with the large $n$ result \cite{KZI}
mentioned above, which predicts a first order transition.
The explanation for the seemingly divergent correlation length
observed in MC simulations could be the following.
For $\beta<\beta_c$ the defects strongly disorder the system
and cause a small correlation length. Above $\beta_c$, however,
the role of the defects decreases rapidly with increasing $\beta$.
As the defects become unimportant the correlation length
approaches that of the $\On$ model.
The numerical simulation of the $\RP2$ model  \cite{KZII}
gave $\beta_c=5.58$ which in the $\O3$ model corresponds to a
correlation length $\xi \sim 10^{15}$! 
A sharp transition or a jump to a huge value is
therefore is not unexpected. This transition is, however, 
associated with the non--universal dynamics of the defects, 
not with the universal continuum limit of the theory.

To establish the equivalence of the $\RPn$ model
(at $\mu=0$) with the $\On$
model in the continuum limit it suffices to show that the defects 
do not play any role in the $\beta\to\infty$ limit.
The defects (or rather pairs of defects) have finite activation energy
which depends on the distance $r$ between the two defects
as $const+{1\over 2}\pi\ln r$. 
The constant contribution coming from the neighborhood of the defects
depends strongly on the actual form of the function $f(u)$
in (\ref{9}),
more precisely on the values of $f(u)$ for small $|u|$, say\footnote{
It is easy to show that around a defect at least one of the four
links has $u^2\le 0.5$.} $u^2<0.5$.
Because the defect pairs have finite
activation energy $E_0$, they are exponentially suppressed by
$\exp(-\beta E_0)$. The subtlety here is that the correlation
volume, $\xi^2(\beta)\propto \exp(4\pi \beta)$ (for $n=3$), is also
exponentially large, and pairs of defects with limited relative 
distances will occur in this volume if their $E_0$ is small 
enough\footnote{For the standard $\RPn$ action the minimal activation
energy is $E_0 ^{min}=2.14$}.
These could be, however, considered as local --- i.e. non--topological
excitations on the scale of $\xi(\beta)$, and we do not expect that
they significantly influence the $\beta\to\infty$ limit.
The argument becomes
even simpler if one changes the form of the action
by pushing up the values of $f(u)$ for $u^2<0.5$ to have $E_0 > 4\pi$
for all defects. In this case the defects
are practically absent in the whole correlation volume
\footnote{Obviously this argument does not apply if
the correlation length becomes infinite already at finite $\beta$ 
as suggested in ref.~\cite{PS}.}.

As a concrete realization of the modified $\RPn$ model we take
\bee
f(u)={1\over 2} (1-u^2)+q\cdot \max(u_0^2-u^2,0).
\label{17}
\ee
Here $q\ge 0$ and we choose $u_0^2=0.8$ for definiteness.
A simple numerical investigation shows that for $q=10$
the activation energy for neighboring defects is $E_0\approx 4\pi$.
(Of course, nothing forbids taking $q=\infty$ --- it will still define
the same continuum theory.)

By similar modifications of the action
it might well be possible to bring the correlation length
down to reasonable values, so that the phase diagram could
be reliably investigated numerically
(also in the mixed $\RPn/\On$ model).
This would imply
that the huge correlation length around the point where the defects
start to condensate for the standard $\RPn$ model
is rather ``accidental''.

\section{The perturbed $\RPn$ model}

Consider the perturbed $\RPn$ model
\bee
\A(S)=\beta_T \sum_{x,\mu} f\left( \SV_x \SV_{x+\mu}\right)
+ \beta_V \sum_{x,\mu} g\left( \SV_x \SV_{x+\mu}\right)
\label{18}
\ee
in the limit $\beta_T \to\infty$, $\beta_V=\,$fixed.
Here $f(u)$ satisfies (\ref{10}), while the perturbation
$g(u)$ can, without loss of generality, be taken to be odd:
\bee
g(-u)=-g(u).
\label{19}
\ee
The action (\ref{1}) is, of course, (up to an irrelevant constant)
a special case.
At $\beta_T\to\infty$ the scalar product
$\SV_x \SV_{x+\mu}$ is forced to be around $+1$ or $-1$, i.e.
$1-(\SV_x \SV_{x+\mu})^2 = O(1/\beta_T)$. 

Let us now assume that $\beta_T$ is large enough or
the form of $f(u)$ is chosen such that the
defects are completely negligible
(as in the example of (\ref{17}) for $q\ge10$).
For configurations with no defects one can introduce the Ising 
variables $\epsilon_x$ 
in a unique way and define the ``true vector'' field $\sig_x$ as in 
(\ref{13}).
Separating the sign of $g(u)$ by
\bee
g(u)=-{\rm sign}(u) g_0(|u|)=
-{\rm sign}(u) \left[ g_0(1)+g'_0(1)(|u|-1)+\ldots \right] ,
\label{20}
\ee
we obtain
\bee
\A(\SV)=\A_V(\sig)+\A_{Ising}(\epsilon)+\A_{int}(\epsilon,\sig),
\label{21}
\ee
where
\bee
\A_V(\sig)\quad=\beta \sum_{x,\mu} f_V(\sig_x \sig_{x+\mu}),
\label{22}
\ee
\bee
\A_{Ising}(\epsilon)=-J  \sum_{x,\mu} \epsilon_x \epsilon_{x+\mu},
\label{23}
\ee
\bee
\A_{int}(\epsilon,\sig)= \sum_{x,\mu} \epsilon_x \epsilon_{x+\mu}
\left[ -g'_0(1)(1-\sig_x \sig_{x+\mu}) + \ldots \right].
\label{24}
\ee
Here $\beta=\beta_T$, $J=\beta_V g_0(1)$ and $f_V(u)$ is
as in (\ref{15}).
Note $1-\sig_x \sig_{x+\mu}=O(1/\beta)$ and hence the interaction
term $\A_{int}(\epsilon,\sig)$ goes effectively to zero as
$\beta\to\infty$. 

Consider first the simple case when $g(u)=-{\rm sign}(u)$, i.e.
$g_0(u)=1$. In this case the two systems decouple exactly while
the specific behavior of the vector and tensor correlation
functions still persists.
Since the correlator $\langle \SV_x \SV_y \rangle$ factorizes:
\bee
\langle \SV_x \SV_y \rangle =
\langle \epsilon_x \epsilon_y \rangle \langle \sig_x \sig_y \rangle ,
\label{25}
\ee
for $J < J_c$ one has 
\bee
m_S=m_{\epsilon}+m_{\sigma} ~~~{\rm and}~~~~  m_T=2 m_{\sigma},
\label{26}
\ee
where the masses are defined through the exponential decay of the 
corresponding correlators.
Although the tensor mass is smaller than twice the vector mass,
$m_T < 2 m_S$, one can not conclude from this that
there is a pole in the tensor channel (in contrast to the pure
$\On$ model), as suggested in ref.~\cite{CEPSII}.
Since both $m_{\sigma}(\beta)$ and $m_{\epsilon}(J)$ go to zero
as $\beta$ and $J$ approach their critical values, the ratio
\bee
 r = {m_T \over m_S}={2 m_{\sigma} \over  m_{\sigma} + m_{\epsilon}}
\label{27}
\ee
can be fixed  at any value $r \in [0,2]$ by properly approaching
the point $(J_c,\infty)$ in the $(J,\beta)$ plane.

For $J>J_c$ the Ising field $\epsilon_x$ develops a non--zero
expectation value hence in this case $m_S=m_{\sigma}$ and
$m_T/m_S=2$.
Note that for finite $\beta$ the phase transition around $J=J_c$
is observed only in the non--local variable $\epsilon_x$ not
in the original variable $\SV_x$ whose correlation length remains
finite at $J=J_c$.

Following the argument in refs.~\cite{CEPSI,CEPSII} one would
conclude that around the point $(J,\beta)=(J_c,\infty)$ one
could define
seemingly inequivalent theories differing in the ratio $m_T/m_S$
\footnote{The masses measured in \cite{CEPSI} are not the true masses,
but those defined through the second
moments; it is however generally believed that the qualitative picture
remains unaltered.}.
Although this is formally true, the corresponding theory is neither
really new nor interesting. In particular, all the tensor
correlation functions are the same as those
in the corresponding pure $\On$ model.

With the choice $g(u)=-u$, i.e. $g_0(u)=|u|$ (as in ~\cite{CEPSI})
the situation is more complicated since there is an interaction
between the two systems. However, as mentioned above, the effective
strength of the interaction goes to zero as $\beta\to\infty$, hence
it might well happen that in the continuum limit one recovers
the previous situation.

Note that the presence or absence of the interaction is not connected
with the behavior of $g(u)$ around $u=+1$ (which is responsible
for the $\On$ continuum limit $\beta_V\to\infty$) but rather with the
difference in behavior around $u=+1$ and $u=-1$.
For example, $g(u)={1\over 2}(1-u^2)+c \theta(-u)$ (not antisymmetrized
in this case) where $c>0$ and $\theta$ is the step function,
is a perfectly acceptable discretization of the $\On$ model
for $\beta_V\to\infty$ and it produces no interaction, $\A_{int}=0$.
On the other hand, $g(u)$ could be chosen to have, say, a local
maximum at $u=+1$ instead of a minimum, which would completely
destroy the $\beta_V\to\infty$ behavior but would still have
the same interaction pattern as for the case $g(u)=-u$.

In this sense, the phenomenon around the point $(J_c,\infty)$ is 
the consequence of perturbing the $\RPn$ model by a term
breaking the local $Z_2$ symmetry, rather than its mixing
with the  $\On$ model.

\section{Some analytic studies of the mixed model}

Let us set $\beta_V=(1-\omega)n/f$ and $\beta_T=\omega n/f$ for
the bare couplings in (\ref{1}). There are various analytic studies
which shed some light on the physics of this model.
Among these are the
ordinary perturbation theory $f\to0$ and the $1/n$ approximation.

\subsection{Bare perturbation theory}

One interesting exercise is to compute the spectrum for a finite
spatial extent $L$.
For the tensor mass $m_T$ to second order
in bare perturbation theory, one finds
\bee
m_T(L) L=f+f^2{1\over n}
\Bigl\{(n-2)R(L/a)+[1+\omega (n+1)]P(L/a)\Bigr\}+O(f^3)
\label{28}
\ee
and to this order the vector mass $m_V$ is given by
\bee
m_V(L) ={(n-1)\over2n}m_T(L).
\label{29}
\ee
In (\ref{28}) the functions
$R,P$ are given by finite sums over lattice momenta.
The relation (\ref{29}) holds before the continuum limit
has been taken (there are no lattice artifacts in the ratio to
this order). Furthermore, the ratio is independent of $\omega$, which is
certainly consistent with notions of universality (the continuum
limit is taken here in finite volumes).
The ratio (\ref{29}) has been shown to hold in the $\On$ model
for small volumes, in the continuum limit to third
order in the renormalized coupling by Floratos and Petcher \cite{FlPe}.
Indeed there, to this order, the mass of the tensor
of rank $k$ is proportional to the eigenvalue of the square Casimir
operator:
\bee
m_k=M k (n+k-2)
\label{30}
\ee
with $M$ independent of $k$. In finite volumes the spectrum is discrete
and there is a finite gap between $m_T$ and $2m_V$; this gap is
expected to close as $L\to\infty$ where a cut develops starting
at $2m_V$.
We have numerically computed the mass
of the tensor as well as that
of the ``true vector" in the $\RPn$ model,
as defined in sect. 2, in small volumes;
the results agreed well with the above formulae.

One can also use (\ref{28}) to determine the ratio of
$\Lambda$-parameters.
For this, it suffices to know
the continuum limit ($a/L \to 0$) behavior of $R,P$:
\bee
R(L/a)\sim {1\over2\pi}
\Bigl\{ \ln(L/a)-\ln(\pi/\sqrt{2})+\gamma_E\Bigr\},
\label{31}
\ee
\bee
P(L/a)\sim {1\over4}
\label{32}
\ee
with $\gamma_E$ Euler's constant.
Denoting $\Lambda(\omega)$ the lattice $\Lambda$-parameter
for a model with given $\omega$,
\bee
{\Lambda(\omega)\over\Lambda(0)}=
\exp \Bigl\{ - {\omega \pi (n+1)\over 2(n-2)}\Bigr\}
\label{33}
\ee
follows, in agreement with the result in ref.~\cite{HaNi}.

Comparing the two theories in infinite volume,
Caracciolo and Pelissetto \cite{CP} also found that the $\RPn$
and the $\On$ models have (apart from the redefinition
of the coupling) the same perturbative expansion.

\subsection{$1/n$ Expansion}

The $1/n$ expansion for the mixed model was to our knowledge
first investigated by Magnoli and Ravanini \cite{MR}.
We disagree, however, with some of their final conclusions.
To discuss this, we first introduce a few formulae.
After introducing
auxiliary fields $A_{\mu}(x),t(x)$ to make the integral quadratic
in the spin-fields and then performing the Gaussian integral, the
partition function in the absence of external fields, takes the form
\bee
Z=const \cdot \int \prod_{x,\mu} {\rm d}A_{\mu}(x)\prod_x {\rm d}t(x)
\exp \Bigl\{ -{n\over2}S_{{\rm eff}}\Bigr\},
\label{34}
\ee
with the effective action
\bee
S_{{\rm eff}}=-{1\over f}\sum_x [s+it(x)] +{\rm tr}\ln {\cal M},
\label{35}
\ee
where ${\cal M}$ is the operator
\bee
{\cal M}=s+it+\sum_{\mu} \left\{-\partial_{\mu}^{*}\partial_{\mu}
+\omega [A_{\mu}\partial_{\mu}^{*}\partial_{\mu}
-(\partial_{\mu}^{*}A_{\mu})(1-\partial_{\mu}^{*})+A_{\mu}^2]\right\}.
\label{36}
\ee
Here $\partial_{\mu}(\partial_{\mu}^{*})$ denote the lattice forward
(backward) derivatives.
One first seeks a stationary point of $S_{{\rm eff}}$ at constant
field configurations $A_{\mu}(x)=1-b$, $t(x)=const$.
Demanding a saddle point
at $t=0$ gives a relation for the constant $s$ in (\ref{35}) as a
function of $b$. With $s$ fixed in this way, one seeks minima of
$S_{{\rm eff}}$ as a function of $b$.

For $\omega=1$ (the pure $\RPn$ model),
the extremal points are shown in fig. 1.
In this case there is a symmetry $b\to-b$
\footnote{Actually, for the pure $\RPn$ case there are
$2^{Volume}$ degenerate minima, due to the local $Z_2$ symmetry.
Elitzur's theorem is not violated by this approximation ---
the local symmetry is not broken spontaneously.}.
Further $b=0$ is an extremal point for all $f$.
For $f<1$, the points $b=0$ are maxima and the non-zero values are
minima. For $f=1^+$, $b=0$ becomes a local (but not absolute)
minimum and two new local maxima develop.
At $f=f_c(1)\approx1.046$ the three minima become degenerate,
and for $f>f_c(1)$ the minimum at $b=0$ is the absolute minimum.
One finds (in the leading order of the $1/n$ expansion) that
at this point the tensor correlation length does not go to infinity:
there is a jump in the order parameter and the phase transition is
thus first order.

\begin{figure}[htb]
\begin{center}
\leavevmode
\epsfxsize=90mm
\epsfbox{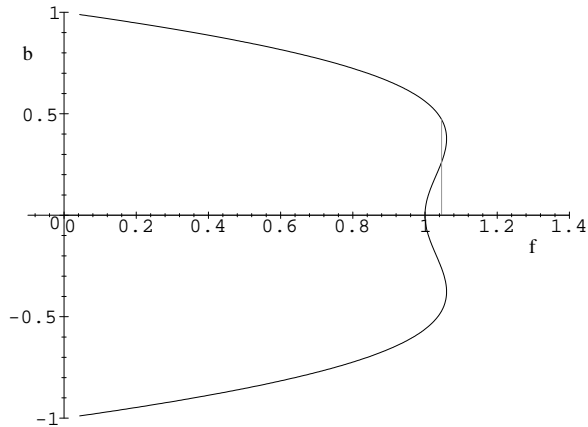}
\vskip 10mm
\end{center}
\caption{The order parameter $b$ in the $1/n$ expansion
as a function of the coupling $f$ for $\omega=1$. There is a jump
at $f_c(1)=1.046$ from a finite value to $b=0$ shown by the dotted line.}
\label{fig:om1000}
\end{figure}
 
For $\omega<1$, the $b\to-b$ symmetry is broken and
the local minimum with $b>0$ is the lowest. For $\omega$ only slightly
less than 1, the situation is as in fig. 2. Here again, at some
$f=f_c(\omega)$ the parameter $b$ undergoes a finite jump.
There is, however, a critical
value of $\omega=\omega_c\approx0.985$ below which the ``S-structure"
in fig. 2 dissolves and there is only one extremal point
for $b>0$ for all values of $f$.
In the $\omega-f$ plane there is thus a first order transition
line which starts at $(1,f_c(1))$, extends only a little way in
the plane and ends at a critical point
${\rm C}=(\omega_c,f_c(\omega_c)\approx 1.075)$.
At C the vector and tensor correlation lengths remain finite.
The transition at C is, however, second order since the specific heat
diverges. The cause of this in the
leading order of the $1/n$ expansion can be traced back to
a development of a singularity in the inverse propagator
of the auxiliary fluctuating $t$-field
\footnote{Note that the $A-$ and $t-$ fields mix and it is
necessary to diagonalize.}
at zero momentum at the critical point. The singularity in the $t-$
propagator seems to remain for higher orders as well. An infinite
correlation length in the energy fluctuations does not contradict a finite
correlation length in the vector and tensor channels; 
in particular, there is no conflict with correlation inequalities. 
These inequalities state that by increasing a ferromagnetic
coupling the system becomes more ordered and the correlation between
any spins increases. 
Although this assumption looks physically quite obvious,
it has not been proven rigorously.
The increase of the correlation function, however, implies
the growing of a correlation length with
increasing ferromagnetic coupling, only when the corresponding
quantity has a vanishing expectation value. 

\begin{figure}[htb]
\begin{center}
\leavevmode
\epsfxsize=90mm
\epsfbox{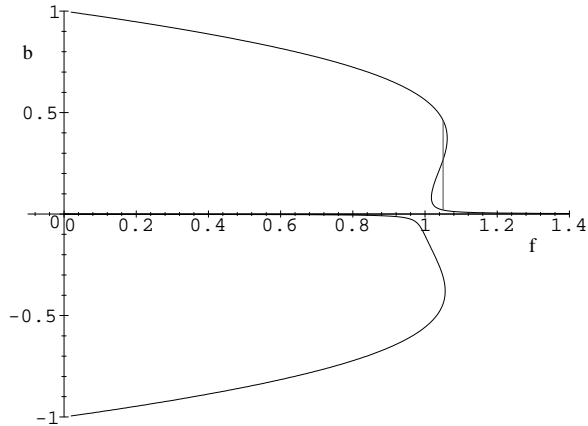}
\end{center}
\caption{The order parameter $b$ as a function of $f$ for $\omega=0.999$.
It still has a finite jump indicated by dotted line.  
At $\omega \ge \omega_c=0.985$ the S--shape dissolves thus the phase
transition disappears.}
\label{fig:om0999}
\end{figure}
 
Thus, a diverging vector (or tensor) correlation length at the endpoint C
would contradict a finite correlation length for large (but finite)
$\beta_V$ (asymptotic freedom) --- on the other hand, a diverging
specific heat at C is not excluded by these considerations.
The above scenario disagrees with that of Magnoli and Ravanini
~\cite{MR} who argue (based on correlation inequalities)
that the second order phase transition at the point C is only
an artifact of the $1/n$ approximation.

Caracciolo, Pelissetto and Sokal \cite{CPS} also discuss the
$\beta_T/N$, $\beta_V$ fixed, $N\to\infty$ limit.
They obtain a result which is equivalent to eq.~(\ref{26}) above
(although their interpretation is different from ours).

\begin{figure}[htb]
\begin{center}
\leavevmode
\epsfxsize=90mm
\epsfbox{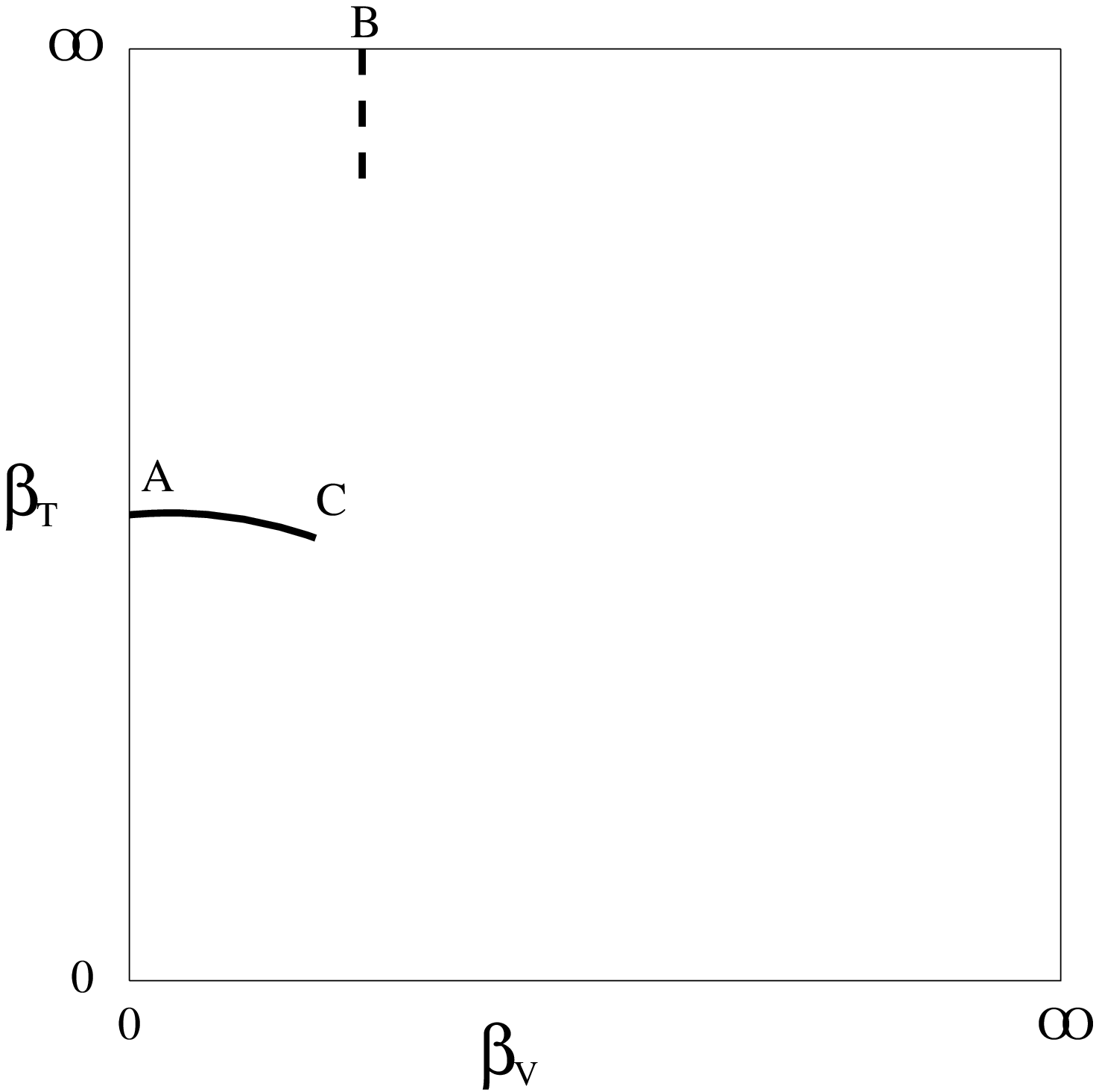}
\end{center}
\caption{The phase diagram for the mixed 
$\RPn$ -- $\On$ model.}
\label{fig:phase_diag}
\end{figure}

In conclusion, it is plausible that the phase diagram, also
for finite $n$, is the ``standard'' one shown in 
fig.~\ref{fig:phase_diag}.
There is a first order transition line starting at the point A
of the $\beta_T$ axis. It ends at the point C where the 
specific heat becomes infinite, but the vector and tensor
correlation lengths remain finite.
In this figure we also indicate the Ising critical point B
discussed in Section~3. The dotted line starting at point B 
is the critical line
of the underlying Ising variable $\epsilon$. This criticality,
however, does not show up in the correlation functions
of the original variable $\SV$.  

{\it Acknowledgements.} 
We thank Peter Hasenfratz and Erhard Seiler for useful discussions
and Alan Sokal for a correspondence.

\eject

\eject

\end{document}